\documentclass[english,aps,nofootinbib,showpacs]{revtex4}
\usepackage[T1]{fontenc}
\usepackage[latin9]{inputenc}
\usepackage{amsmath}
\usepackage{graphicx}
\usepackage{amssymb}
\usepackage{esint}

\makeatletter

\newcommand{\lyxdot}{.}

\@ifundefined{textcolor}{}
{%
 \definecolor{BLACK}{gray}{0}
 \definecolor{WHITE}{gray}{1}
 \definecolor{RED}{rgb}{1,0,0}
 \definecolor{GREEN}{rgb}{0,1,0}
 \definecolor{BLUE}{rgb}{0,0,1}
 \definecolor{CYAN}{cmyk}{1,0,0,0}
 \definecolor{MAGENTA}{cmyk}{0,1,0,0}
 \definecolor{YELLOW}{cmyk}{0,0,1,0}
 }

\usepackage{amsfonts}
\providecommand{\U}[1]{\protect\rule{.1in}{.1in}}

\makeatother

\usepackage{babel}

\begin{document}

\title{Statistics of resonances in one-dimensional disordered systems}

\author{E. Gurevich}

\author{B. Shapiro}

\affiliation{Technion - Israel Institute of Technology, Haifa, Israel, 32000}


\begin{abstract}
The paper is devoted to the problem of resonances in one-dimensional
disordered systems. Some of the previous results are reviewed and
a number of new ones is presented. These results pertain to different
models (continuous as well as lattice) and various regimes of disorder
and coupling strength. In particular, a close connection between resonances
and the Wigner delay time is pointed out and used to obtain information
on the resonance statistics.
\end{abstract}

\date{\today}

\maketitle

\section{Introduction}

\begin{figure}
\begin{centering}
\includegraphics[clip,height=1.6in]{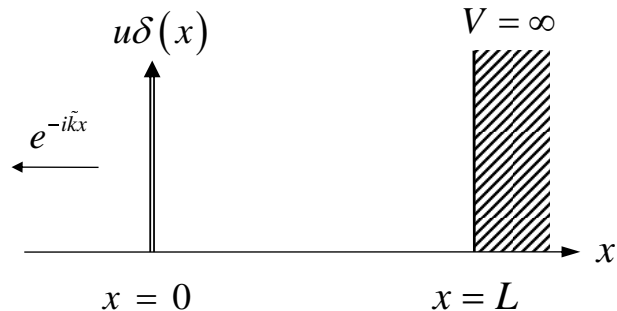}
\par\end{centering}

\caption{An illustration to the problem of resonances in a potential comprised
of a $\delta$-function barrier and a hard wall.}
\label{Fig: wall-delta}
\end{figure}

The problem of resonances, also referred to as metastable or quasi-stationary
states \cite{Landau-V3}, goes back to the early days of quantum mechanics
\cite{Gamov-28}. A simple example of resonances \cite{Kumar-book}
is provided by a potential depicted in Fig. \ref{Fig: wall-delta}.
There is a wall, $V=\infty$, for $x\geq L$ and a potential $V\left(x\right)=u\delta\left(x\right)$.
For $u\rightarrow\infty$, a particle of mass $m$ has bound states
at energies $\frac{1}{2m}\left(\frac{\pi\hbar n}{L}\right)^{2}$ {[}$n=1,2,\ldots${]}.
For any finite $u$ the spectrum becomes continuous. However, the
strictly stationary states which existed at $u=\infty$ do leave a
trace in the continuum and turn into resonances. They correspond to
poles of the scattering matrix $S\left(E\right)$ on the unphysical
sheet of the complex energy plane \cite{Landau-V3},\cite{Baz-book}.
An alternative, more direct approach to the problem of resonances
amounts to solving the stationary Schrödinger equation with the boundary
condition of an outgoing wave only \cite{Landau-V3},\cite{Gamov-28}.
Thus for the potential in Fig. \ref{Fig: wall-delta} one has to solve
the equation\begin{equation}
-\frac{d^{2}\psi}{dx^{2}}+\alpha\delta\left(x\right)\psi=\tilde{k}^{2}\psi\qquad\left(\alpha=\frac{2mu}{\hbar^{2}}\right)\label{Schrod-delta}\end{equation}
 with the boundary condition $\psi\left(x=L\right)=0$ and the outgoing
wave condition $\psi\left(x\right)=e^{-i\tilde{k}x}$ for $x<0$.
The latter condition makes the problem non-hermitian: the eigenvalues
for $\tilde{k}$, and for the corresponding \textquotedbl{}energies\textquotedbl{}
$\tilde{E}=\hbar^{2}\tilde{k}^{2}/2m$, will be generally complex.

The solutions of Eq. (\ref{Schrod-delta}) is\begin{equation}
\psi\left(x\right)=\left\{ \begin{array}{ll}
A\sin\tilde{k}(x-L),\quad & 0<x<L\\
e^{-i\tilde{k}x}, & x<0\end{array}\right..\label{Sol_Schrod-delta}\end{equation}
 Matching the function and its derivative at $x=0$ results in\begin{equation}
\left(1-i\gamma\tilde{k}L\right)\tan\tilde{k}L=-\gamma\tilde{k}L,\label{Match_Schrod-delta}\end{equation}
 where $\gamma=\left(\alpha L\right)^{-1}\ll1$. For $\gamma=0$ one
recovers the bound states, $k_{n}L=\pi n$. For small $\gamma$ the
solutions of (\ref{Match_Schrod-delta}) are obtained by iteration:\begin{equation}
\tilde{k}_{n}L=\pi n\left(1-\gamma+\gamma^{2}\right)-i\left(\pi n\right)^{2}\gamma^{2}+O\left(\gamma^{3}\right).\end{equation}
 One can immediately write down the \textquotedbl{}eigenenergies\textquotedbl{},
$\tilde{E}_{n}=\hbar^{2}\tilde{k}_{n}^{2}/2m=E_{n}-\frac{i}{2}\Gamma_{n}$.
The real part, $E_{n}$, gives the position of the resonance on the
energy axis, whereas $\Gamma_{n}$ determines the resonance width.
For $n$ not too large, namely, $n\ll\gamma^{-1}$, the resonances
are sharp, i.e., their width is much smaller than their spacing on
the energy axis. This simple example demonstrates how true bound states
in a closed system ($u\rightarrow\infty$) turn into resonances, when
the system is opened to the outside world (finite $u$, i.e. non-zero
coupling constant $\gamma$).

\begin{figure}
\begin{centering}
\includegraphics[clip,height=1.6in]{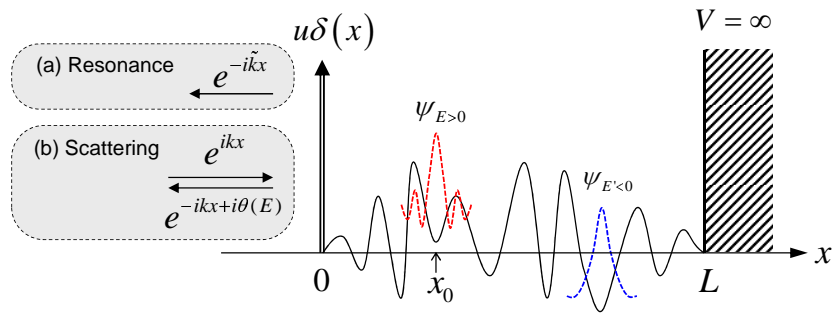}
\par\end{centering}

\caption{Schematic illustration of a 1D disordered system with one end ($x=L$)
closed, and the other one ($x=0$) coupled to the lead through the
$\delta$-function barrier. The resonance problem corresponds to (a):
the outgoing wave condition is imposed and complex values of $\tilde{k}$
are found. The standard scattering problem is described in the set-up
(b), where a particle with energy $E=\hbar^{2}k^{2}/2m$ is impinging
on the system.}

\centering{}\label{Fig: disorder}
\end{figure}

Open quantum systems can be described in terms of an effective, non-Hermitian
Hamiltonian whose complex eigenvalues give the position of the resonances
in the complex energy plane (in addition, there might be real eigenvalues
which correspond to the bound states). Such non-Hermitian Hamiltonians
have been used for a long time in scattering theory, including scattering
in disordered and chaotic systems \cite{Weidenmuller-69},\cite{Datta},\cite{Dittes-00},\cite{Kottos}.
There is a considerable amount of work on resonances in disordered
potentials \cite{Terraneo},\cite{Pinheiro},\cite{Titov-00},\cite{Weiss-06},\cite{Kottos-LDT-02},\cite{Kunz-Sh-06},\cite{Kunz-Sh-08},\cite{Feinberg},\cite{Feinberg-09}.
An example of one-dimensional random potential is depicted in Fig.
\ref{Fig: disorder}. The potential $V\left(x\right)$ is zero for
$x\leq0$ and it is infinite for $x\geq L$. In the interval $0<x<L$,
$V\left(x\right)$ is a random function of $x$, with zero mean and
some well defined statistical properties. There is also a barrier
$u\delta(x)$ at $x=0$ which allows to tune the coupling strength
to the external world. For $u\rightarrow\infty$ (closed system) all
states are localized within the system. Two such localized wave function
are schematically shown in the figure: $\psi_{E}\left(x\right)$ is
a state of positive energy, localized far away from the boundary $x=0$,
i.e. its localization center $x_{0}$ is much larger than the localization
length $\xi$. The function $\psi_{E^{\prime}}\left(x\right)$ corresponds
to a negative energy state, which is localized essentially in a single
deep potential well. When $u$ is made finite the localized state
$\psi_{E}\left(x\right)$ will turn into a narrow resonance, with
a width $\Gamma$ proportional to $\exp\left(-2x_{0}/\xi\right)$,
while the state $\psi_{E^{\prime}}\left(x\right)$ will remain a true
bound state. A theory of resonances in disordered chains should consider
the statistical ensemble of all possible realizations of $V\left(x\right)$
and produce the probability distribution $P\left(\Gamma\right)$.

A quantity closely related to the resonance width is the Wigner delay
time \cite{Wigner-55},\cite{Nussenzveig-02} which is a measure of
the time spent by the particle in the scattering region and is defined
as the energy derivative of the scattering phase shift. For the single-channel
scattering, as presented in the setup (b) in Fig. \ref{Fig: disorder},
the solution of the scattering problem amounts to finding the phase
$\theta\left(E\right)$ of the reflected wave, $e^{-ikx+i\theta\left(E\right)}$,
due to the incident wave $e^{ikx}$. The corresponding Wigner delay
time is defined as \begin{equation}
\tau\left(E\right)=\hbar\frac{d\theta\left(E\right)}{dE}.\label{Wigner DT}\end{equation}
 For a disordered system, $\theta\left(E\right)$ and $\tau\left(E\right)$
are random quantities, characterized by the joint distribution $P_{E,L}\left(\theta,\tau\right)$
over the ensemble of realizations. There exists a large body of work
on the statistics of delay times for the scattering on disordered
and chaotic systems \cite{Kumar-89},\cite{Kumar-00},\cite{Jayannavar-98},\cite{Osipov-Kot-00},\cite{Comtet-97},\cite{Comtet-99},\cite{Fyodorov-97}.
In the presence of a sharp, well isolated resonance $\tilde{E}_{n}=E_{n}-\frac{i}{2}\Gamma_{n}$,
delay time at the energy $E$ close to $E_{n}$ is given approximately
by \cite{Nussenzveig-02} \begin{equation}
\tau\left(E\right)\approx\hbar\frac{\Gamma_{n}}{\left(E_{n}-E\right)^{2}+\Gamma_{n}^{2}/4},\end{equation}
 which demonstrates the intimate relation between the resonance width
and the delay time. Below (section IV) we obtain a relation between
the delay time and the resonance width distributions which is exact
in the limit of weak coupling to the lead, and which enables us to
obtain information about resonances based on the existing knowledge
of the delay time statistics.

Statistics of resonances and of delay times (or the closely related
\textquotedbl{}dwell times\textquotedbl{}) are of great interest in
the physics of disordered media. For instance, in a disordered conductor
the current carriers can be trapped for a long time, which lead to
long tails in the decay of an electric current \cite{Mirlin-00}.
Although our discussion is limited to \textquotedbl{}matter waves\textquotedbl{},
obeying the Schrödinger equation, similar phenomena occur for electromagnetic
waves as well. When a wave is injected into a random dielectric medium,
it can spend there a very long time, before escaping from the sample.
This phenomenon of long delay times has been extensively studied in
experiments \cite{Genak-05}. Resonances and long escape times might
be also relevant to the phenomenon of \textquotedbl{}random lasing\textquotedbl{},
when an active random dielectric medium without any prefabricated
cavities, exhibits lasing above some excitation threshold \cite{Cao-03}.

The organization of the paper is as follows. In section \ref{Sect: TBM}
we introduce a tight binding model and, following \cite{Kunz-Sh-06},\cite{Kunz-Sh-08}
derive the effective non-Hermitian Hamiltonian for the resonance problem.
Section \ref{Sect: PT} is devoted to the case when coupling between
the disordered chain and the external lead is weak. Both the tight
binding model and a random continuous potential are treated. In section
\ref{Sect: Res2DT} a relation between the distributions of resonances
and delay times is obtained, and used for studying properties of the
resonances under various conditions (weak and strong coupling, finite
and infinite chain). Section \ref{Sect: Strong DO} specializes to
the case of strong disorder, for the tight binding model, using the
locator expansion technique.

\section{A tight binding model for resonances and its effective Hamiltonian\label{Sect: TBM}}

\begin{figure}
\begin{centering}
\includegraphics[clip,height=1.5in]{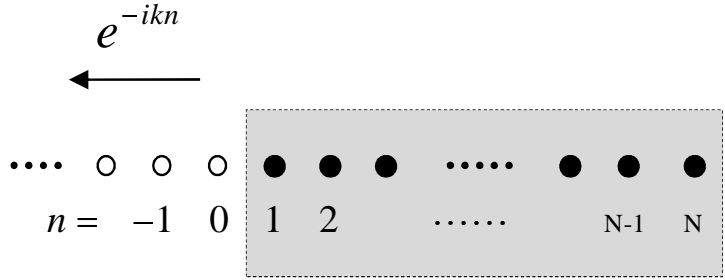}
\par\end{centering}

\caption{Resonance problem for one dimensional TBM.}

\centering{}\label{Fig: TBM}
\end{figure}

Along with the model of a continuous random potential, described in
the Introduction, we also consider the tight binding (Anderson) model
(TBM) depicted in Fig. \ref{Fig: TBM}. Black dotes, labeled by $n=1,2,...N$,
designate sites in the chain. Each site is assigned a site energy
$\epsilon_{n}$ chosen from some distribution $q(\epsilon)$. The
energies on different sites are independent of each other. Open circles,
labeled by $n=0,-1,-2,\ldots$, represent the perfect semi-infinite
lead to which the chain is coupled. The lead simulates the free space
outside the chain. All nearest neighbor sites of the chain are coupled
to each other by a hopping amplitude $t$, and the same is true for
all nearest neighbor sites of the lead. The only exception to this
rule is the pair $n=\left(0,1\right)$ which provides coupling between
the chain and the lead. The hopping amplitude for this pair is taken
to be equal $t^{\prime}$. This allows us to tune the coupling from
$t^{\prime}=0$ (closed chain) to $t^{\prime}=t$ (perfect coupling).
The Schrödinger equation for the entire system (chain + lead) is a
set of coupled equations:\begin{align}
-t\psi_{n+1}-t\psi_{n-1} & =\tilde{E}\psi_{n}\quad\quad(n<0)\label{chain_1}\\
-t\psi_{-1}-t^{\prime}\psi_{1} & =\tilde{E}\psi_{0}\quad\quad(n=0)\label{chain_2}\\
-t\psi_{2}-t^{\prime}\psi_{0}+\epsilon_{1}\psi_{1} & =\tilde{E}\psi_{1}\quad\quad(n=1)\label{chain_3}\\
-t\psi_{n+1}-t\psi_{n-1}+\epsilon_{n}\psi_{n} & =\tilde{E}\psi_{n}\quad\quad(1<n\leq N).\label{chain_4}\end{align}
 with the Dirichlet boundary condition $\psi_{N+1}=0$. Eqs. (\ref{chain_1})-(\ref{chain_4})
are to be solved subjected to the boundary condition of an outgoing
wave in the lead, i.e. $\psi_{n}\propto\exp(-i\tilde{k}n)$, for $n\leq0$,
with $\operatorname{Re}\tilde{k}>0$ (the wave propagates from right
to left). The complex wave vector $\tilde{k}$ is related to $\tilde{E}$
by $\tilde{E}=-2t\cos\tilde{k}$. The complex solutions $\tilde{E}_{\alpha}=E_{\alpha}-\frac{i}{2}\Gamma_{\alpha}$
of Eqs. (\ref{chain_1})-(\ref{chain_4}) yield the width of the resonances,
as well as their position along the energy axes $E$.

As has been explained in the Introduction, the condition of an outgoing
wave makes the problem a non-Hermitian one. In particular, for the
tight binding model (Fig. \ref{Fig: TBM}) one can derive an explicit
expression for an effective non-Hermitian Hamiltonian whose eigenvalues
correspond to the resonances, in addition to the possible bound states.
Using the plane wave solution $\psi_{n}\propto\exp(-i\tilde{k}n)$
in the lead $(n<1)$, it is straightforward to eliminate from Eqs.
(\ref{chain_1})-(\ref{chain_4}) all $\psi_{n}$'s with $n<1$ (for
details see \cite{Datta}), thus reducing the problem to a system
of equations for the amplitudes $\psi_{n}$ on the sites of the disordered
chain alone ($n=1,2,...N$): \begin{equation}
-t\psi_{n+1}-t\psi_{n-1}+\tilde{\epsilon}_{n}\psi_{n}=\tilde{E}\psi_{n}\quad\quad(n=1,2,\ldots,N)\,\label{chain1}\end{equation}
 with the boundary conditions $\psi_{0}=\psi_{N+1}=0$. Here $\tilde{\epsilon}_{n}=\epsilon_{n}$
for $n=2,3,....$, but not for $n=1$. This end site is assigned a
complex energy \begin{equation}
\tilde{\epsilon}_{1}=\epsilon_{1}-t\eta e^{i\tilde{k}},\label{epsilon}\end{equation}
 where the parameter $\eta=\left(t^{\prime}/t\right)^{2}$ describes
the coupling strength to the outside world. Thus, the effective non-Hermitian
Hamiltonian $\tilde{H}$, defined in (\ref{chain1}), differs from
the Hermitian Hamiltonian, $H$, of the corresponding closed system
(i.e., with $\eta=0$) only by the complex correction to the energy
of the first site (the only site coupled directly to the lead), i.e.,
\begin{equation}
\tilde{H}=H-t\eta e^{i\tilde{k}}P,\label{H_eff}\end{equation}
 where $P$ is the projection on site $n=1$. Note that the effective
Hamiltonian $\tilde{H}$ depends, via $\tilde{k}$, on $\tilde{E}$.
Therefore Eq. (\ref{chain1}) does not constitute a standard eigenvalue
problem and the eigenvalues of $\tilde{H}$ have to be determined
self-consistently. We denote the complex variable $\tilde{E}\equiv z=E-\frac{i}{2}\Gamma$.
It is shown in \cite{Kunz-Sh-08} that the resonances $z_{\alpha}$,
in the complex $z$-plane, correspond to the roots (with $\operatorname{Im}z_{\alpha}<0$)
of the equation \begin{equation}
z-\epsilon_{1}-S_{1}(z)+\eta te^{i\tilde{k}(z)}=0,\label{F}\end{equation}
 where $S_{1}(z)$ is the self-energy for site $n=1$ and $\tilde{k}(z)$
is related to $z$ by $z=-2t\cos\tilde{k}$ (in \cite{Kunz-Sh-08}
the variable $z$ was measured in units of $t$).

\section{Treating the coupling term in $\tilde{H}$ as perturbation\label{Sect: PT}}

When the coupling to the lead is weak ($\eta\ll1$), the resonances
can be obtained as small corrections to the eigenvalues of the closed
system. For the tight binding effective Hamiltonian $\tilde{H}$,
Eq. (\ref{H_eff}), first order perturbation theory with respect to
the coupling term $-t\eta e^{i\tilde{k}}P$ gives\begin{equation}
\tilde{E}_{\alpha}=e_{\alpha}-\eta t\psi_{\alpha}^{2}\left(1\right)e^{ik_{\alpha}}\equiv E_{\alpha}-\frac{i}{2}\Gamma_{\alpha},\end{equation}
 where $e_{\alpha}$ is the energy of the unperturbed eigenstate $\psi_{\alpha}$
{[}the former notation $\psi_{n}$ has been changed into $\psi_{\alpha}\left(n\right)$
where subscript $\alpha$ labels the eigenstates{]}, related to $k_{\alpha}$
by\begin{equation}
e_{\alpha}=-2t\cos k_{\alpha}\Rightarrow e^{ik_{\alpha}}=-\frac{e_{\alpha}}{2t}+i\sqrt{1-\frac{e_{\alpha}^{2}}{4t^{2}}}.\end{equation}
 Thus, the resonance width is\begin{equation}
\Gamma_{\alpha}=2\eta\psi_{\alpha}^{2}\left(1\right)\sin k_{\alpha}=2\eta\psi_{\alpha}^{2}\left(1\right)\sqrt{t^{2}-\frac{1}{4}e_{\alpha}^{2}}.\label{Gamma_TBM_PT}\end{equation}
 In addition to the imaginary correction, $-i\Gamma/2$, there is
also a real-valued correction, i.e. $E_{\alpha}=e_{\alpha}+\eta\psi_{\alpha}^{2}\left(1\right)e_{\alpha}/2$.
This small energy shift on the real axis is of no interest. Note that
the resonances exist only for $\left\vert e_{\alpha}\right\vert <2t$,
i.e. within the band of the lead. For energies outside the band only
bound states exist (real $\tilde{E}_{\alpha}$).

An expression analogous to (\ref{Gamma_TBM_PT}) is obtained also
for the continuous case depicted in Fig.\textbf{ }\ref{Fig: disorder},
either as a continuum limit of (\ref{Gamma_TBM_PT}) or by direct
application of the perturbation theory. In the latter approach, matching
the internal solution to the outgoing wave in the lead, one obtains\begin{equation}
\frac{\psi\left(x=0^{+},\tilde{k}\right)}{\psi^{\prime}\left(x=0^{+},\tilde{k}\right)}=\frac{1}{i\tilde{k}-\frac{2mu}{\hbar^{2}}}.\label{Secular eq}\end{equation}
 Here $\psi\left(x,\tilde{k}\right)$ is the solution in the interval
$0<x<L$ for the energy $E=\frac{\hbar^{2}\tilde{k}^{2}}{2m}$, which
satisfies the closed-end boundary condition $\psi\left(x=L,\tilde{k}\right)=0$
plus the condition of the outgoing wave for $x\,<0$. For $u\rightarrow\infty$,
Eq. (\ref{Secular eq}) gives the spectrum and the eigenstates of
the closed system, satisfying zero boundary condition $\psi_{\alpha}\left(0,k_{\alpha}\right)=0$.
For weak coupling to the lead, $g\equiv\frac{2mu}{\hbar^{2}k_{\alpha}}\gg1$,
perturbative expansion of the above secular equation in powers of
$g^{-1}$, yields\begin{equation}
\Gamma_{\alpha}=4\frac{\hbar^{2}k_{\alpha}^{2}}{2m}\left[\frac{\hbar^{2}}{2mu}\right]^{2}\frac{\psi^{\prime}\left(0,k_{\alpha}\right)}{\psi_{k}\left(0,k_{\alpha}\right)},\label{Gamma_PT_cont}\end{equation}
 where $\psi_{k}\left(x,k\right)=\frac{\partial}{\partial k}\psi\left(x,k\right)$.
Then, employing the identity (see, e.g., Refs. \cite{Comtet-99},\cite{Nussenzveig-02})\begin{equation}
\left\vert \psi\right\vert ^{2}=\frac{\hbar^{2}}{2m}\frac{d}{dx}\left(\frac{d\psi^{\ast}}{dx}\frac{d\psi}{dE}-\psi^{\ast}\frac{d^{2}\psi}{dxdE}\right),\label{WF_Identity}\end{equation}
 the resonance width is expressed as\begin{equation}
\Gamma_{\alpha}=\frac{\hbar^{2}k_{\alpha}}{2m}\left[\frac{\hbar^{2}k_{\alpha}}{mu}\right]^{2}\frac{\left\vert \psi_{\alpha}^{\prime}\left(0,k_{\alpha}\right)\right\vert ^{2}}{2k_{\alpha}^{2}},\label{Gamma_PT_cont_fin}\end{equation}
 where $\psi_{\alpha}$ is the normalized to unity eigenfunction of
the closed system with the eigenenergy $E_{\alpha}=\frac{\hbar^{2}k_{\alpha}^{2}}{2m}$.
This expression is consistent with the exact effective Hamiltonian
for the continuous open systems derived in Refs.\ \cite{Pichugin-01},\cite{Savin-03}.

Thus, both in the continuum and in TBM, for weak coupling to the lead
there is one-to-one correspondence between the resonances and the
eigenstates of the closed system, and the resonance width is related
to the tail of the corresponding eigenstate at the boundary.

Certain simplifications occur in the limit of a semi-infinite chain.
The TBM in the $N\rightarrow\infty$ limit has been studied in \cite{Kunz-Sh-08},
where the small-$\Gamma$ asymptotics for the density of resonances
(DOR) has been rigorously derived in the weak coupling limit ($\eta\ll1$).
DOR in the $E,\Gamma$-plane, for a given realization of the disorder,
is given by\begin{equation}
\rho(E,\Gamma)=\sum_{\alpha}\delta(E-E_{\alpha})\delta(\Gamma-\Gamma_{\alpha}),\label{DOR_def}\end{equation}
 where $z_{\alpha}=E_{\alpha}-\frac{i}{2}\Gamma_{\alpha}$ are solutions
of (\ref{F}). For any $\Gamma\neq0$, this expression for DOR has
a well defined $N\rightarrow\infty$ limit and no division of the
sum by $N$ is necessary, - in contrast to the usual case of the density
of states (on the real axis) for a Hermitian problem. Note that the
probability distribution of resonance width $P(\Gamma)$ (for some
fixed $E$) does not have a well defined $N\rightarrow\infty$ limit
and it approaches $\delta(\Gamma)$. (Indeed, for a semi-infinite
chain an eigenstate will be localized, with probability $1$, at an
infinite distance from the open end and, thus, will be ignorant about
the coupling to the external world.) Thus, the appropriate quantity
to look at for a semi-infinite chain is the DOR, rather than the probability
distribution of resonance width. This subtle point is discussed in
some detail in \cite{Kunz-Sh-06}.

Although the general considerations in \cite{Kunz-Sh-08} pertain
to any coupling strength $\eta$, specific results for the average
DOR where obtained only in the weak coupling limit, where the width
of all resonances becomes proportional to $\eta$. The small-$\Gamma$
asymptotics for the average DOR $\left\langle \rho(E,\Gamma)\right\rangle $
is \cite{Kunz-Sh-08} (wherein the result is written in terms of some
rescaled variables): \begin{equation}
\left\langle \rho(E,\Gamma)\right\rangle ={\frac{\nu\left(E\right)\xi\left(E\right)}{2\Gamma}},\label{rho}\end{equation}
 where $\nu(E)$ and $\xi(E)$ are, respectively, the usual density
of states (on the real energy axis) and the localization length for
an infinite disordered chain. Angular brackets denote averaging over
the ensemble of all random realizations. This asymptotic ($1/\Gamma$)-
behavior is universal, in the sense that it holds for any degree of
disorder and for any $-2t<E<2t$.

The $1/\Gamma$-asymptotics can be understood with the help of a simple
intuitive argument which, in somewhat different versions, has appeared
in \cite{Terraneo},\cite{Pinheiro},\cite{Titov-00},\cite{Weiss-06},\cite{Comtet-99}.
The essence of the argument is that narrow resonances stem from states
localized far away from the open boundary, say, at distance $x$.
Such states will have an exponentially small tail at the boundary,
proportional to $e^{-x/\xi}$, and the corresponding resonances will
be exponentially narrow, $\Gamma\sim e^{-2x/\xi}$. The $1/\Gamma$-
behavior then immediately follows from the assumption that the localization
centers, $x$, are uniformly distributed in space.

One should keep in mind that, for a long but finite chain of $N$
sites, the $\left(1/\Gamma\right)$-tail will be cut off at very small
$\Gamma$ of the order of $\exp\left(-2N/\xi\right)$. The extremely
narrow resonances with $\Gamma\ll\exp\left(-2N/\xi\right)$ originate
from states localized in the vicinity of the closed-end site $n=N$
and they should be treated separately (see below).

\section{Relation between distributions of resonances and delay times\label{Sect: Res2DT}}

The rigorous asymptotic result of the previous section, Eq. (\ref{rho}),
was obtained for a \emph{semi-infinite} chain \emph{weakly} coupled
to an external lead. Things get more complicated if these restrictions
are relaxed. In particular, the simple relation between the resonance
width and the behavior of the corresponding eigenstate of the closed
system {[}Eqs. (\ref{Gamma_TBM_PT}) and (\ref{Gamma_PT_cont_fin}){]}
breaks down when the coupling between the system and the lead becomes
strong. In this section we discuss systems of finite size $L$ and
beyond weak coupling limit.

In the Introduction we have mentioned the problem of the delay time
$\tau\left(E,L\right)$ and the corresponding phase shift $\theta\left(E,L\right)$,
for a particle of energy $E$ impinging on a random chain of length
$L$. We designate by $P_{E,L}\left(\theta,\tau\right)$ the joint
probability distribution of $\theta$ and $\tau$ \emph{for perfect
coupling} to the lead ($\eta=1$, or $u=0$) and relate this distribution
to the average DOR $\left\langle \rho\left(E,\Gamma\right)\right\rangle $.
Such relation is useful because it enables us to \textquotedbl{}transfer\textquotedbl{}
the existing knowledge of the time delay in disordered chains \cite{Kumar-89},\cite{Kumar-00},\cite{Jayannavar-98},\cite{Osipov-Kot-00},\cite{Comtet-97},\cite{Comtet-99}
into the field of resonances. To this end we introduce the quantity
\begin{equation}
\Phi_{\alpha}=\frac{1}{2}\left\vert \frac{\psi_{\alpha}^{\prime}\left(0\right)}{k}\right\vert ^{2},\label{Y_alf_def}\end{equation}
 where $\psi_{\alpha}\left(x\right)$ is a normalized eigenfunction
of the closed system satisfying the boundary conditions $\psi\left(0\right)=\psi\left(L\right)=0$.
The average density of points $\left\{ E_{\alpha},\Phi_{\alpha}\right\} $
in the $\left(E,\Phi\right)$-plane is\begin{equation}
\left\langle \tilde{\rho}\left(E,\Phi\right)\right\rangle =\left\langle \sum_{\alpha}\delta\left(E-E_{\alpha}\right)\delta\left(\Phi-\Phi_{\alpha}\right)\right\rangle .\label{DOY}\end{equation}
 Although $\left\langle \tilde{\rho}\left(E,\Phi\right)\right\rangle $
is defined in terms of eigenvalues and eigenfunctions of the closed
system, it can be related to the distribution $P_{E,L}\left(\theta,\tau\right)$
which describes scattering properties of the corresponding open system.
The relation stems from the fact that for $\theta\left(E\right)=\pi$
the scattering wave function vanishes at $x=0$, so that the eigenvalues
$E_{\alpha}$ are given by zeros of the function $\theta\left(E\right)-\pi$.
This observation results in the identity\begin{equation}
\delta\left(\theta\left(E\right)-\pi\right)=\sum_{\alpha}\left[\frac{d\theta\left(E\right)}{dE}\right]^{-1}\delta\left(E-E_{\alpha}\right)=\frac{\hbar}{\tau\left(E\right)}\sum_{\alpha}\delta\left(E-E_{\alpha}\right).\end{equation}

A generalization of this identity involves, in addition to the eigenvalues
$E_{\alpha}$, also the eigenfunction-related quantity $\Phi_{\alpha}$,
(\ref{Y_alf_def}), and it reads \cite{Gurevich-PhD}\begin{equation}
\sum_{\alpha}\delta\left(E-E_{\alpha}\right)\delta\left(\Phi-\Phi_{\alpha}\right)=\frac{4}{\hbar v_{g}^{2}\Phi^{3}}\delta\left(\theta\left(E\right)-\pi\right)\delta\left(\tau\left(E\right)-\frac{2}{v_{g}\Phi}\right),\end{equation}
 where $v_{g}=\hbar^{-1}\frac{dE}{dk}$ is the group velocity in the
lead. This identity, upon averaging over the distribution $P_{E,L}\left(\theta,\tau\right)$
and using (\ref{DOY}), yields the required relation between the quantities
characterizing the open and the closed system: \begin{equation}
\left\langle \tilde{\rho}\left(E,\Phi\right)\right\rangle =\frac{4}{\hbar v_{g}^{2}\Phi^{3}}P_{E,L}\left(\theta=\pi,\tau=\frac{2}{v_{g}\Phi}\right).\label{P(Y)2P(DT)_cont}\end{equation}
 This expression holds for arbitrary $L$ and has a well defined $L\rightarrow\infty$
limit {[}cf. the discussion after Eq. (\ref{DOR_def}){]}. Let us
note that Eq. (\ref{P(Y)2P(DT)_cont}) constitutes the strictly one-dimensional
counterpart of the similar relations derived in Ref. \cite{Fyodorov-05}
for the one-channel scattering from a higher-dimensional system. The
results in Ref. \cite{Fyodorov-05} \ were obtained within the nonlinear
sigma model and, thus, do not include the strictly 1D case discussed
here.

Equations (\ref{Y_alf_def}),(\ref{P(Y)2P(DT)_cont}) correspond to
the continuous model. A completely similar treatment for the TBM yields
precisely the same relation (\ref{P(Y)2P(DT)_cont}) {[}with $\hbar\equiv1${]},
but with $\Phi_{\alpha}$ redefined as \begin{equation}
\Phi_{\alpha}=\frac{\psi_{\alpha}^{2}\left(1\right)}{2\sin^{2}k},\label{P(Y)2P(DT)_TBM}\end{equation}
 and the group velocity in the lead given by $v_{g}=\frac{dE}{dk}=2t\sin k$.

The relation (\ref{P(Y)2P(DT)_cont}) is rather general. It holds
for an arbitrary $L$, for any degree of disorder, and it is applicable
to lattice models as well as to continuous ones. However, to employ
this relation for the resonance statistics problem one more step is
needed, namely, a relation between $\Phi_{\alpha}$ and $\Gamma_{\alpha}$.
For the weak coupling case such a relation has been derived in the
previous section for both the TBM {[}Eq. (\ref{Gamma_TBM_PT}){]}
and the continuous potential (\ref{Gamma_PT_cont_fin}). The two expressions
can be unified into a single formula\begin{equation}
\Gamma_{\alpha}=T\frac{\hbar v_{g}}{2}\Phi_{\alpha},\quad\left(T\ll1\right)\label{Gamma2Y}\end{equation}
 where $T$ is the transmission coefficient through the potential
barrier separating the lead from the chain. The latter is realized
by a $\delta$-function potential in the continuum or by the weak
hopping link $t^{\prime}$ in the TBM, as described previously, so
that\begin{equation}
T=\left[\begin{array}{ll}
\frac{1}{1+\left(mu/\hbar^{2}k\right)^{2}},\smallskip & \text{continuous model}\\
\frac{4\eta\sin^{2}k}{\left(1-\eta\right)^{2}+4\eta\sin^{2}k}, & \text{TBM}\end{array}\right..\end{equation}
 Note that the linear relation (\ref{Gamma2Y}) between $\Gamma_{\alpha}$
and $\Phi_{\alpha}$ is valid only if $T$ is small (weak coupling).
With the help of (\ref{Gamma2Y}) one can map the density $\left\langle \tilde{\rho}\left(E,\Phi\right)\right\rangle $
in the $\left(E,\Phi\right)$-plane, Eq. (\ref{P(Y)2P(DT)_cont}),
onto the average DOR in the $\left(E,\Gamma\right)$-plane:\begin{equation}
\left\langle \rho\left(E,\Gamma\right)\right\rangle =\frac{\hbar T^{2}}{\Gamma^{3}}P_{E,L}\left(\theta=\pi,\tau=T\frac{\hbar}{\Gamma}\right).\label{DOR2PDT_weak_coupl}\end{equation}
 This formula relates the average DOR to the delay time statistics.
For a weak Gaussian white noise disorder and $L\gg\xi$ the distribution
$P_{E,L}\left(\theta,\tau\right)$ does not depend on $\theta$ and
has the following form \cite{Comtet-97}:\begin{equation}
P_{E,L}\left(\theta,\tau\right)=\frac{\tau_{0}}{2\pi\tau^{2}}e^{-\tau_{0}/\tau}+\frac{1}{\pi^{2}\tau}e^{-\tau_{0}/2\tau}\int_{0}^{\infty}dse^{-L\left(1+s^{2}\right)/2\xi}\frac{s}{1+s^{2}}\sinh\frac{\pi s}{2}W_{1,is/2}\left(\frac{\tau_{0}}{\tau}\right),\label{P(z,L)_exact_Comte}\end{equation}
 where $\tau_{0}=\xi/v_{g}$ and $W_{1,is/2}$ is the Whittaker function
(the same result is obtained for the weak correlated disorder \cite{Gurevich-PhD}).
Expression (\ref{P(z,L)_exact_Comte}), via (\ref{DOR2PDT_weak_coupl}),
immediately yields the corresponding DOR.

In the limit $L/\xi\rightarrow\infty$, $\tau$ fixed, Eq. (\ref{P(z,L)_exact_Comte})
reduces to\begin{equation}
P_{E,\infty}\left(\theta,\tau\right)=\frac{1}{2\pi}\frac{\tau_{0}}{\tau^{2}}e^{-\tau_{0}/\tau},\label{P(DT)_infinit}\end{equation}
 so that\begin{equation}
\left\langle \rho\left(E,\Gamma\right)\right\rangle =\frac{\nu_{0}\xi}{2\Gamma}e^{-\Gamma/\Gamma_{0}},\label{rho_2}\end{equation}
 where $\nu_{0}=\left(\pi\hbar v_{g}\right)^{-1}$ is the density
of states in the lead per unit length and\begin{equation}
\Gamma_{0}=T\frac{v_{g}\hbar}{\xi}.\label{Gamm_0_def}\end{equation}
 Eq. (\ref{rho_2}) coincides with the former result (\ref{rho})
for $\Gamma/\Gamma_{0}\ll1$ and, in addition, gives an exponential
suppression of the resonance density for $\Gamma/\Gamma_{0}>1$ (the
exact density of states $\nu$ in (\ref{rho}) reduces to $\nu_{0}$
in the weak disorder limit).

\begin{figure}[ptb]
\begin{centering}
\includegraphics[width=3.0801in,height=2.3155in]{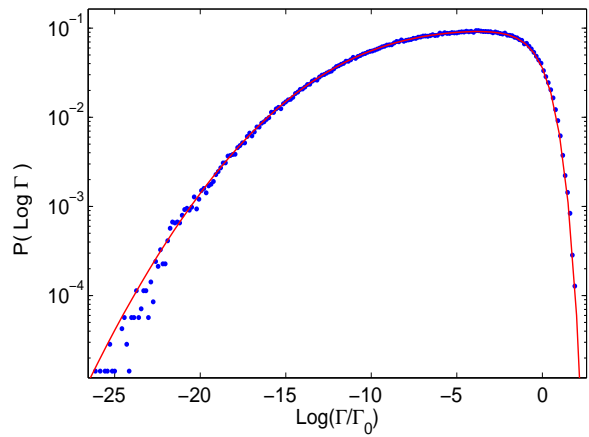}\caption{Weak coupling to the lead. The numerical simulation (dots) was done
for the TBM (\ref{chain_1})-(\ref{chain_4}) with $t=1$, $E=-1.9$,
$\eta=0.001$ and system length $N=401$ sites. The uncorrelated disorder
with a box distribution was implemented, which produced localization
length $\xi=77.6$ sites. The analytical curve was calculated according
to Eqs. (\ref{DOR2PDT_weak_coupl}) and (\ref{P(z,L)_exact_Comte}).}

\par\end{centering}

\centering{}\label{Fig: weak coupl}
\end{figure}

For finite size chain ($L/\xi\approx5$) and weak coupling ($T\approx.0004$)
the distribution $P\left(\log\Gamma\right)=\left(\Gamma/\nu L\right)\left\langle \rho\left(E,\Gamma\right)\right\rangle $,
calculated from (\ref{DOR2PDT_weak_coupl}),(\ref{P(z,L)_exact_Comte})
is presented in Fig. \ref{Fig: weak coupl} (solid line). For comparison,
a numerical Monte-Carlo simulation was performed for the TBM at the
energy close to the unperturbed band edge (dots in Fig. \ref{Fig: weak coupl}).
The agreement is quite good. Let us discuss the example in Fig. \ref{Fig: weak coupl}
in more detail. First, the exponential factor $\exp\left[-\Gamma/\Gamma_{0}\right]$,
which suppresses the large-$\Gamma$ probability, is present in both
the semi-infinite, Eq. (\ref{rho_2}), and finite-$L$, Eq. (\ref{P(z,L)_exact_Comte}),
case. For $\Gamma$'s smaller than the characteristic value $\Gamma_{0}$,
one can distinguish two regimes. In the intermediate regime, $e^{-2L/\xi}\ll\Gamma/\Gamma_{0}<1$,
the behavior $P\left(\log\Gamma\right)\approx const$ {[}i.e. $\left\langle \rho\left(E,\Gamma\right)\right\rangle \sim1/\Gamma$,
Eq. (\ref{rho_2}){]} is valid, since the opposite closed boundary
of the system has not yet come into play. On the contrary, the regime
of very narrow resonances, $\Gamma\ll\Gamma_{0}e^{-2L/\xi}$, is strongly
affected by the boundary $x=L$. These resonances are associated with
the eigenstates localized close to this boundary and are described
by the nearly log-normal tail of the distribution.

Although the regime of the narrow resonances, $\Gamma\ll\Gamma_{0}e^{-2L/\xi}$
is contained in the analytical expressions (\ref{DOR2PDT_weak_coupl}),(\ref{P(z,L)_exact_Comte}),
it is worthwhile to give an independent, more direct derivation. Let
us recall that the localization length $\xi$ is defined for an \emph{infinite}
system and, in this limit, it is a self-averaging quantity. In a long
but finite size chain ($L\gg\xi$) the localizaton length, or more
precisely its inverse (the Lyapunov exponent $\lambda$) is a fluctuating
quantity with nearly a Gaussian distribution (see e.g. \cite{Comtet-99}
and references therein)\begin{equation}
P_{\lambda}\left(\lambda;L\right)=\sqrt{\frac{L\xi}{2\pi}}e^{-L\xi\left(\lambda-\xi^{-1}\right)^{2}/2}.\label{P(lambda)}\end{equation}
 The tail of the extremely small $\Gamma$'s is related to the eigenstates
localized near the closed end of the system, $x=L$, for which $\Phi_{\alpha}\approx\Phi_{0}e^{-2L\lambda}$,
where the pre-factor $\Phi_{0}\sim\xi^{-1}$ is of minor importance.
Then, neglecting the pre-exponential factor, the probability for $\Phi_{\alpha}\ll\Phi_{0}e^{-2L/\xi}$
decays like\begin{equation}
P\left(\Phi\right)\propto\exp\left[-\frac{\left(\ln\left(\Phi/\Phi_{0}\right)+2L/\xi\right)^{2}}{8L/\xi}\right].\end{equation}
 Using (\ref{Gamma2Y}), one obtains the $\log$-normal cutoff of
the DOR\begin{equation}
\left\langle \rho\left(E,\Gamma\right)\right\rangle \propto\exp\left[-\frac{\left(\ln\left(\Gamma/\Gamma_{0}\right)+2L/\xi\right)^{2}}{8L/\xi}\right],\quad\frac{\Gamma}{\Gamma_{0}}\ll e^{-2L/\xi}\label{DOR_LN_WeakDO}\end{equation}
 Similar cutoffs for the delay time and the average DOR have been
derived in Refs. \cite{Comtet-99} and \cite{Titov-00} respectively.

So far the discussion was limited to the weak coupling case, when
a simple relation between $\Phi_{\alpha}$ and $\Gamma_{\alpha}$
{[}Eq. (\ref{Gamma2Y}){]} could be rigorously derived. When the coupling
parameter $T$ increases and approaches unity, the relation (\ref{Gamma2Y})
ceases to be quantitatively accurate and turns into an order of magnitude
estimate $\Gamma_{\alpha}\sim\hbar v_{g}\Phi_{\alpha}$. This relation
is physically reasonable for narrow, isolated resonances. Such resonances
stem from the eigenstates (of the closed system), which are localized
far away from the open boundary $x=0$, and their width is much smaller
than the mean level spacing. One can then trace a particular resonance,
i.e. its width $\Gamma$ as a function of the increasing coupling
strength $T$, without worrying about other resonances. It is therefore
intuitively clear that the small-$T$ result, Eq. (\ref{Gamma2Y}),
can be qualitatively extrapolated up to the perfect coupling limit
$T=1$.

One can support the above argument by a more elaborated analysis.
Consider the formal solution of the one-channel scattering problem
at energy $E$ close to a \emph{narrow isolated} resonance $z_{r}=E_{r}-\frac{i}{2}\Gamma$.
For small $E-E_{r}$, using general analytical properties of the scattering
amplitude in the complex energy plane, the solution in the lead can
be expanded as (see, e.g., Ref. \cite{Razavy})\begin{equation}
\psi\left(x,E\right)=a\left(E-E_{r}+\frac{i}{2}\Gamma\right)e^{ikx}+a^{\ast}\left(E-E_{r}-\frac{i}{2}\Gamma\right)e^{-ikx},\quad x<0,\label{Scat_sol_formal}\end{equation}
 where, by identity (\ref{WF_Identity}), the complex constant $a$
satisfies (up to small corrections) \begin{equation}
\left\vert a\right\vert ^{2}=\frac{1}{\hbar v_{g}\Gamma}\int_{0}^{L}\left\vert \psi\left(x\right)\right\vert ^{2}dx.\label{a^2}\end{equation}
 Energy $E_{\alpha}$ at which $\psi\left(x=0,E_{\alpha}\right)=0$
is the eigenenergy of the closed system, and by (\ref{Scat_sol_formal})\begin{equation}
E_{\alpha}-E_{r}=\frac{\operatorname{Im}a}{2\operatorname{Re}a}\Gamma.\label{EigEn2Res}\end{equation}
 For the corresponding eigenfunction of the closed system, using (\ref{Scat_sol_formal}),(\ref{a^2})
in the definition (\ref{Y_alf_def}), one obtains\begin{equation}
\Phi_{\alpha}\approx\frac{\Gamma}{2\hbar v_{g}}\left[\left(\frac{\operatorname{Im}a}{\operatorname{Re}a}\right)^{2}+1\right]\geq\frac{\Gamma}{2\hbar v_{g}}.\label{Y_alf_near_res}\end{equation}
 Both (\ref{EigEn2Res}) and (\ref{Y_alf_near_res}) are meaningful
as long as $\left(E_{\alpha}-E_{r}\right)\lesssim\Gamma$, i.e., $\left(\frac{\operatorname{Im}a}{\operatorname{Re}a}\right)^{2}\lesssim1$,
since otherwise the linear expansion (\ref{Scat_sol_formal}) is not
valid and higher orders should be included. With this reservation,
Eq. (\ref{Y_alf_near_res}) relates narrow isolated resonances to
the well localized eigenstates of the closed system. However, contrary
to (\ref{Gamma2Y}), relation (\ref{Y_alf_near_res}) is not deterministic,
since it depends on the phase of $a$ (which is random for weak disorder).
Replacing the unknown coefficient $\left[\left(\frac{\operatorname{Im}a}{\operatorname{Re}a}\right)^{2}+1\right]$
by a phenomenological constant $\beta^{-1}$ leads to\begin{equation}
\Gamma_{\alpha}=2\beta\hbar v_{g}\Phi_{\alpha}.\label{Gamma2Y_strong}\end{equation}

With the relation (\ref{Gamma2Y_strong}) at hand, all the steps done
for the weak coupling can be repeated, and Eqs. (\ref{DOR2PDT_weak_coupl}),(\ref{P(z,L)_exact_Comte})
and (\ref{rho_2}) apply with the transmission coefficient $T$ replaced
by $4\beta$ and the characteristic value $\Gamma_{0}$ {[}Eq. (\ref{Gamm_0_def}){]}
redefined as \begin{equation}
\Gamma_{0}=4\beta\frac{v_{g}\hbar}{\xi}.\label{Gamm_0_ideal}\end{equation}
 In the present case, however, the DOR obtained from Eqs. (\ref{DOR2PDT_weak_coupl}),(\ref{P(z,L)_exact_Comte})
is valid only for $\Gamma\ll\Gamma_{0}$, since otherwise the isolated
resonance approximation implied in the above argument is not applicable.

\begin{figure}
\begin{minipage}[b]{8cm}%
\centering{ \includegraphics[width=3.2943in,height=2.4093in]{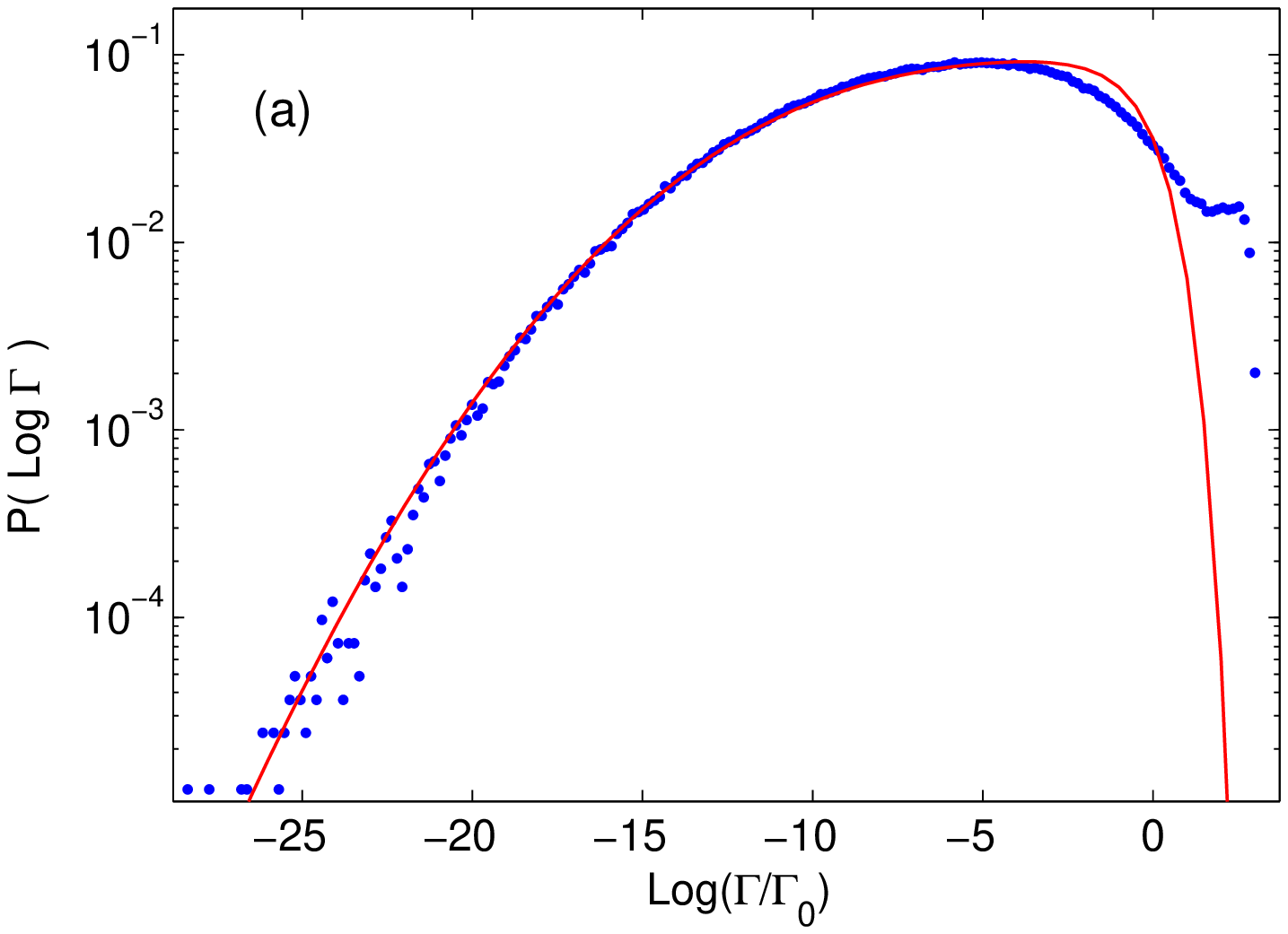}}%
\end{minipage}\hspace{0.8cm} %
\begin{minipage}[b]{8cm}%
\centering{\includegraphics[width=3.2893in,height=2.4093in]{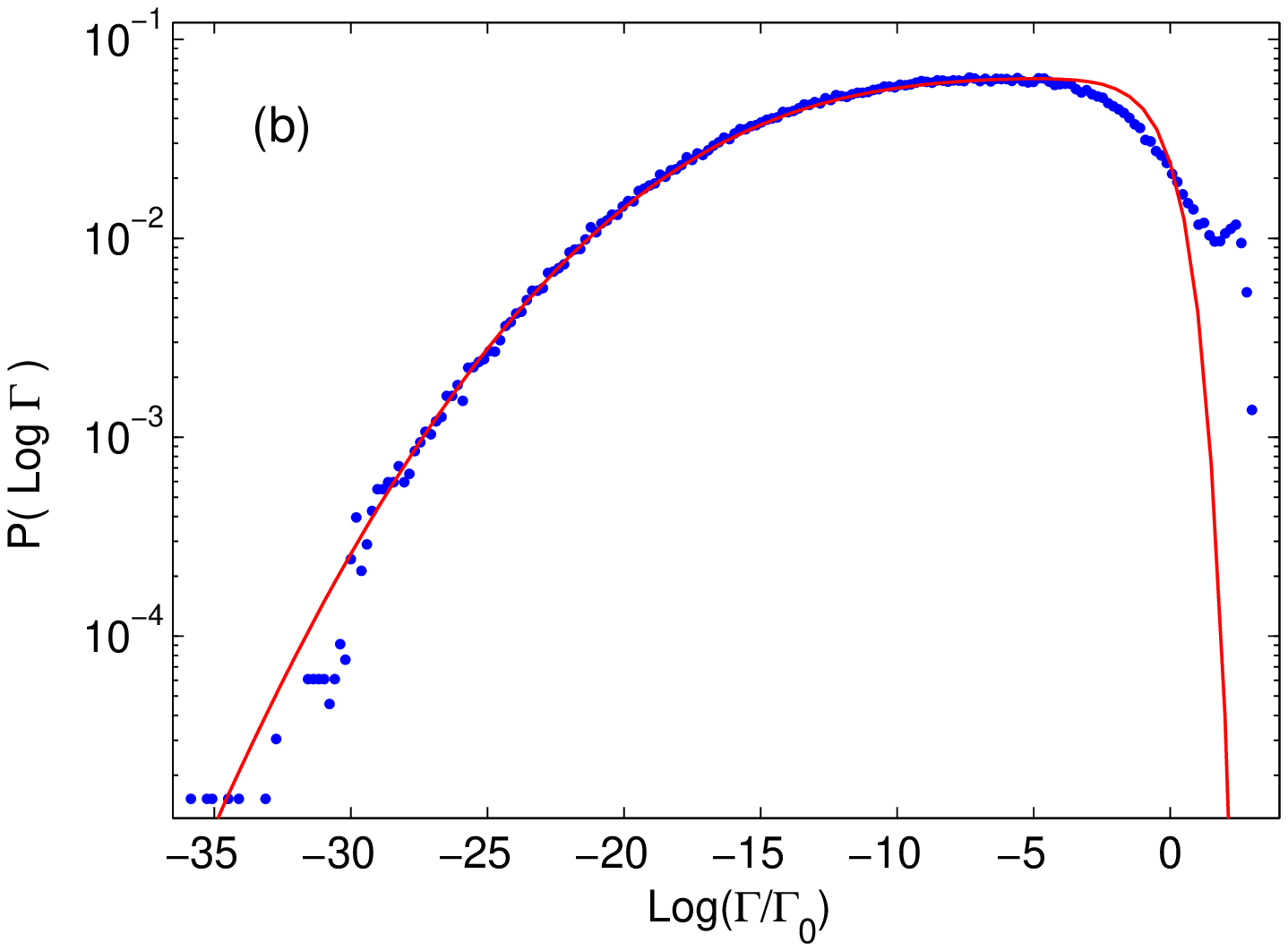}}%
\end{minipage}\caption{Resonance width distribution for the perfect coupling $T=1$. The
analytical curve is calculcated according to Eqs. (\ref{DOR2PDT_weak_coupl}),(\ref{P(z,L)_exact_Comte})
with $\Gamma_{0}$ given by (\ref{Gamm_0_ideal}). The numerical simulation
was done as explained in Fig. \ref{Fig: weak coupl}, with the TBM
parameters $t=1$, $E=-1.9$, $\eta=1$ and the uncorrelated disorder
resulting in the localization length $\xi=77.6$ sites. The system
length is (a) $N=401$ sites and (b) $N=601$ sites. The scatter of
the numerical data for the extremely small values of $\Gamma/\Gamma_{0}$
occurs because of the insufficient numerical statistics. }

\label{Fig: Ideal_coupl} 
\end{figure}

The above approximation was compared to the numerical simulation for
the perfect coupling to the lead, Fig. \ref{Fig: Ideal_coupl}. In
both cases shown in Fig. \ref{Fig: Ideal_coupl}, $L/\xi\approx\allowbreak5$
and $L/\xi\approx\allowbreak8$, the same fitting value $\beta\approx\allowbreak0.68$
was used. As expected, a good agreement between the numerical simulation
(dots) and the analytical result (solid line) is obtained only for
$\Gamma<\Gamma_{0}$ (the deviation for the extremely small $\Gamma$'s
is due to the numerical under-sampling).

\section{Strong Disorder\label{Sect: Strong DO}}

In this section we consider the case of strong disorder, when the
hopping amplitude $t$ is much smaller than the characteristic width
$W$ of the site energy distribution $q(\epsilon)$. For a semi-infinite
chain the problem was considered in \cite{Kunz-Sh-06}, making use
of a recursion relation for the self-energy. Here we employ the locator
expansion, i.e. perturbation theory in $t$, which is the appropriate
tool for strong disorder \cite{Anderson-58},\cite{Ziman}. Our treatment
is not restricted to a semi-infinite chain and, in particular, we
address the question of the cutoff of the ($1/\Gamma$)-tail in a
chain of large but finite $N$. Furthermore, no restriction on the
coupling strength $\eta$ is imposed in our treatment.

\begin{figure}
\begin{centering}
\includegraphics[clip,height=0.5in]{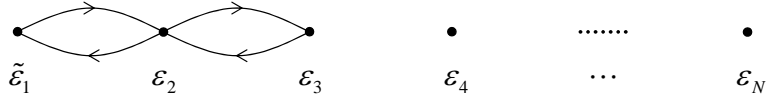}
\par\end{centering}

\caption{Calculation of the self-energy in the locator expansion.}
\label{FIg: locator}
\end{figure}

For $t=0$ the Hamiltonian (\ref{H_eff}) corresponds to uncoupled
sites and its eigenvalues coincide with the site energies $\epsilon_{j}$
($j=1,2,....N$). When $t$ is switched on, some of these \textquotedbl{}unperturbed\textquotedbl{}
eigenvalues acquire a complex correction, due to the last term in
(\ref{H_eff}), and thus describe resonances. Our purpose is to find
the imaginary part of this correction, in the leading order in $t$.
(The small correction to the real part, $\epsilon_{j}$, introduces
an unessential shift on the real axis of the complex energy plane
and will be ignored). We designate the complex energy $\tilde{E}$
by $z=E-\frac{i}{2}\Gamma$ and look for the solutions, $z_{j}$,
of Eq. (\ref{F}), which we rewrite as \begin{equation}
z-\tilde{\epsilon}_{1}-S_{1}(z)=0,\label{F1}\end{equation}
 with \begin{equation}
\tilde{\epsilon}_{1}=\epsilon_{1}-\eta te^{i\tilde{k}(z)}.\label{F2}\end{equation}
 In order to see the mechanism by which the unperturbed solutions,
$z_{j}^{(0)}=\epsilon_{j}$, acquire an imaginary correction, we employ
the locator expansion for the self-energy $S_{1}(z)$. It can be represented
diagrammatically as a sum over all paths which start at site $1$
and return to this site only once \cite{Anderson-58},\cite{Ziman}.
An example of such a path is drawn in Fig. \ref{FIg: locator}. This
path goes from site $1$ to $2$, proceeds from $2$ to $3$ and returns
back to $1$ . This path contributes to $S_{1}$ a term $tg_{2}tg_{3}tg_{2}t$,
where $g_{n}=(z-\epsilon_{n})^{-1}$ is the Green's function (the
locator) for an isolated site $n$. Thus, the general rule is that
to a line connecting a pair of sites one assigns the number $t$,
while to a site $n$ the corresponding locator is assigned. By inspecting
Eq. (\ref{F1}) it becomes clear that an imaginary correction to the
unperturbed solution $z_{j}^{(0)}$ is produced by paths, in the $S_{1}$
- expansion, which connect site $1$ to site $j$. Indeed, site $j$
has no direct knowledge about the connection to the outside world:
this information must be transmitted to it from site $1$, via all
intermediate sites. To leading order, it suffices to keep the shortest
path. For site $3$ this is the path in Fig \ref{FIg: locator}. Generalization
to an arbitrary site $j$ is obvious and results in a path of $(j-1)$
loops which brings in a factor $t^{2(j-1)}$. This path produces the
imaginary part of $z_{j}$, which is calculated from Eq.(\ref{F1}):
\begin{equation}
\operatorname{Im}z_{j}=t^{2(j-1)}\prod_{k=2}^{j-1}\frac{1}{(\epsilon_{j}-\epsilon_{k})^{2}}\operatorname{Im}\frac{1}{\epsilon_{j}-\tilde{\epsilon}_{1}}.\label{zj}\end{equation}
 Since only the leading term (in powers of $t$) is kept, we have
replaced in all the locators $z$ by $z_{j}^{(0)}=\epsilon_{j}$.
For the same reason, $\tilde{k}(z)$ in the expression (\ref{F2})
can be replaced by $k(E)$. From the relation $E=-2t\cos k$ it follows
that \begin{equation}
\tilde{\epsilon}_{1}=\epsilon_{1}+t\eta\left(\frac{E}{2t}-i\sqrt{1-\frac{E^{2}}{4t^{2}}}\right).\label{F3}\end{equation}
 Note that the imaginary part in (\ref{F3}) exists only for $|E|<2t$,
i.e. only bound states in this energy interval (in a closed chain)
turn into resonances upon coupling the chain to the lead (the same
energy interval has already been identified in Sec. \ref{Sect: PT}).
Eigenstates beyond this energy interval remain strictly bound states.
Substituting (\ref{F3}) into (\ref{zj}) and, again, keeping only
leading terms in $t$, one finally obtains: \begin{equation}
-\operatorname{Im}z_{j}\equiv\frac{\Gamma_{j}}{2}=t^{2(j-1)}\eta\sqrt{t^{2}-\frac{E^{2}}{4}}\prod_{k=1}^{j-1}\frac{1}{\left(\epsilon_{j}-\epsilon_{k}\right)^{2}}.\label{zj1}\end{equation}

The DOR in the $\left(E,\Gamma\right)$-plane is given by \begin{equation}
\rho(E,\Gamma)=\sum_{j=2}^{N}\delta\left(E-\epsilon_{j}\right)\delta\left(\Gamma-\Gamma_{j}\right).\label{dor}\end{equation}
 Since the small shift of the eigenvalues along the real axis is of
no interest, we have set $E_{j}=\epsilon_{j}$ in Eq. (\ref{dor}).
For a fixed $j$, resonance width $\Gamma_{j}$ depends on the energies
of all previous sites, $k=1,2,\ldots,j-1$, but not on $\epsilon_{j}$.
Therefore, the two $\delta$-functions in (\ref{dor}) are statistically
independent, so that upon averaging\begin{equation}
\left\langle \rho(E,\Gamma)\right\rangle =q\left(E\right)\left\langle \sum_{j=2}^{N}\delta\left(\Gamma-\Gamma_{j}\right)\right\rangle ,\end{equation}
 where, in the strong disorder limit, the site energy distribution
function $q\left(E\right)$ coincides with the density of states per
site in the closed system. To avoid cluttering the notation we set
$E=0$ (middle of the band) and $\eta=1$ (perfect coupling). (Extension
to arbitrary $E$ and $\eta$ requires some obvious minor modifications.)

For this case\begin{equation}
\left\langle \rho(E=0,\Gamma)\right\rangle =q\left(0\right)\sum_{j=2}^{N}\left\langle \delta\left(\Gamma-2t\prod_{k=1}^{j-1}\frac{t^{2}}{\epsilon_{k}^{2}}\right)\right\rangle .\label{dor_x0}\end{equation}
 It is convenient to define a random variable\begin{equation}
A_{j}=\ln\prod_{k=1}^{j-1}\frac{t^{2}}{\epsilon_{k}^{2}}=-2\sum_{k=1}^{j-1}\ln\frac{\left\vert \epsilon_{k}\right\vert }{t}.\label{A_j}\end{equation}
 This is a sum of independent random variables with the average value
$\left\langle \ln\frac{\left\vert \epsilon\right\vert }{t}\right\rangle =\int d\epsilon q\left(\epsilon\right)\ln\frac{\left\vert \epsilon\right\vert }{t}\equiv\alpha$
and variance $\left\langle \ln^{2}\frac{\left\vert \epsilon\right\vert }{t}\right\rangle -\alpha^{2}\equiv\beta$.
For instance, for the Anderson model, when $\epsilon$ is uniformly
distributed within a window $-\frac{tW}{2}<\epsilon<\frac{tW}{2}$,
one has $\alpha=\left(\ln\frac{W}{2}-1\right)$ and $\beta=1$. Note
that, in the strong disorder limit, $\alpha$ coincides with the inverse
localization length (the Lyapunov exponent) \cite{Thouless-72}.

Since $N$ is a very large number, most of terms in (\ref{dor_x0})
correspond to large $j$, so that $A_{j}$ has a Gaussian distribution,
$P_{j}\left(A\right)$, with the average value $\left\langle A_{j}\right\rangle =-2\left(j-1\right)\alpha$
and variance $\left\langle \Delta A_{j}^{2}\right\rangle =4\beta j$,
i.e.\begin{equation}
P_{j}\left(A\right)=\frac{1}{\sqrt{8\pi\beta j}}\exp\left[-\frac{\left(A+2j\alpha\right)^{2}}{8j\beta}\right],\end{equation}
 where $\left(j-1\right)$ was replaced by $j$. Eq. (\ref{dor_x0})
then yields\begin{align}
\left\langle \rho(E=0,\Gamma)\right\rangle  & =q\left(0\right)\sum_{j=2}^{N}\int\delta\left(\Gamma-2te^{A}\right)P_{j}\left(A\right)dA\nonumber \\
 & =\frac{q\left(0\right)}{\Gamma}\sum_{j=2}^{N}\frac{1}{\sqrt{8\pi\beta j}}\exp\left[-\frac{\left(\ln\left(\Gamma/2t\right)+2j/\xi\right)^{2}}{8j\beta}\right],\label{dor_x0_2}\end{align}
 where $\xi=1/\alpha$ is the localization length in the middle of
the band ($E=0$). The lower limit of summation, $j=2$, should not
be taken literally and it is of no importance, since for small resonance
width $\Gamma$ the sum is dominated by large-$j$ terms.

For narrow (but not too narrow) resonances, when $1\ll-\ln\left(\Gamma/2t\right)\ll N/\xi$,
the sum is dominated by terms with $j$ near $j_{0}\approx-\left(\xi/2\right)\ln\left(\Gamma/2t\right)\gg1$.
Then, the sum in (\ref{dor_x0_2}) can be approximated by an integral
and\begin{equation}
\left\langle \rho(E=0,\Gamma)\right\rangle \approx\frac{q\left(0\right)\xi\left(0\right)}{2\Gamma},\end{equation}
 in agreement with the universal result in Eq. (\ref{rho}). This
$1/\Gamma$ behavior is cut off sharply for very narrow resonances,
such that $-\ln\left(\Gamma/2t\right)\gg N/\xi$. These resonances
stem from states which are localized in the vicinity of the sample
boundary at $j=N$. The sum (\ref{dor_x0_2}) is then dominated by
the last term, i.e.\begin{equation}
\left\langle \rho(E=0,\Gamma)\right\rangle \approx\frac{q\left(0\right)}{\Gamma}\frac{1}{\sqrt{8\pi\beta N}}\exp\left[-\frac{\left(\ln\left(\Gamma/2t\right)+2N/\xi\right)^{2}}{8N\beta}\right],\label{dor_x0_Log-N}\end{equation}
 i.e. for $-\ln\left(\Gamma/2t\right)>2N/\xi$ the DOR rapidly (faster
than any power of $\Gamma$) approaches zero with decreasing $\Gamma$.
This kind of $\log$-normal tails are well known in the theory of
disordered electronic systems \cite{Mirlin-00}.

It is instructive to compare the strong disorder result, Eq. (\ref{dor_x0_Log-N}),
with the expression (\ref{DOR_LN_WeakDO}) which was derived in the
opposite case of weak disorder. The main difference between the two
expressions, besides the fact that in (\ref{DOR_LN_WeakDO}) the pre-exponential
factor has not been written down, is that the exponent in (\ref{DOR_LN_WeakDO})
contains the single parameter $L/\xi$, whereas (\ref{dor_x0_Log-N})
depends in addition on the parameter $\beta/\alpha$ {[}indeed, $8N\beta$
can be written as $8\left(N/\xi\right)\left(\beta/\alpha\right)${]}.
The parameter $\beta/\alpha$ is a non-universal number which depends,
for instance, on the chosen distribution for the site energies, $q\left(\varepsilon\right)$.
The same situation is well known to occur in the study of the transmission
coefficient $T$ through a disordered chain of length $L$. The distribution
of $\ln T$ is Gaussian. If the disorder is weak, then there is a
universal relation between the mean and the variance of $\ln T$ (single
parameter scaling). On the other hand, for strong disorder the two
become independent of one another (two parameter scaling) \cite{Shapiro-88}.

\section{Conclusion}

Statistics of resonances in disordered one-dimensional chains is a
formidable problem which does not easily lend itself to a rigorous
analysis. In this paper we have reviewed some of the existing results
and have extended them in various directions. We consider both a continuous
random potential and the tight binding lattice model, and we tackle
a variety of different cases, differing by size $L$ of the chain,
by strength of the disorder or by coupling strength between the system
and the external world. There is no efficient universal method for
treating the problem in its full generality. Different techniques
turn out to be appropriate in different regimes. In particular, we
presented in some detail the method of locator expansion, most suitable
for the strongly disordered lattice model. On the other hand, for
weak disorder we were able to use some known rigorous results for
the Wigner delay time problem to obtain information on resonance statistics.

BS is indebted to H. Kunz for previous collaboration on the subject.
We acknowledge useful discussions with A. Comtet, J. Feinberg and
C. Texier.

\end{document}